\documentstyle[12pt]{article}
\def\NPB#1#2#3{Nucl. Phys.  {\ B#1}  (19#2)  #3}
\def\PLB#1#2#3{Phys. Lett. {\ B#1} (19#2) #3}

\def\PRD#1#2#3{Phys. Rev. {\ C#1} (19#2) #3}
\def\PRL#1#2#3{Phys. Rev. Lett. {\#1} (19#2) #3}

\def\ARNP#1#2#3{Ann. Rev. Nucl. Part. Sci. {\#1} (19#2) #3}

\def\la{\lambda}
\def\f{\frac}
\def\ne{\nu_{e}}
\def\nm{\nu_{\mu}}

\def\sq{\sqrt{2}}

\def\beq{\begin{equation}}
\def\eeq{\end{equation}}
\def\bea{\begin{eqnarray}}
\def\eea{\end{eqnarray}}
\newcommand{\newc}{\newcommand}
\newc{\sm}{Standard Model}
\newc{\smd}{Standard Model}
\newc{\barr}{\begin{eqnarray}}
 \newc{\earr}{\end{eqnarray}}
\newc{\dac}{discrete anomaly cancellation }
\newc{\mup}{``$\mu$'' problem }
\newc{\eps}{\epsilon}
\def\gappeq{\mathrel{\rlap {\raise.5ex\hbox{$>$}}
{\lower.5ex\hbox{$\sim$}}}}

\def\lappeq{\mathrel{\rlap{\raise.5ex\hbox{$<$}}
{\lower.5ex\hbox{$\sim$}}}}

\begin{document}

\begin{center}
{\bf TEXTURES FOR
NEUTRINO MASS MATRICES IN GAUGE THEORIES
} \\
\vspace*{1.0cm}
{\bf S. Lola$^{a}$} and
{\bf J. D. Vergados\footnote{Lecture given by J.D.Vergados, Erice
1997, published in Progress in 
Particle and Nuclear Physics 40 (1998) 71.
}}
$^{b}$
\end{center}
\begin{center}
\begin{tabular}{l}
$^{a}$
{\small CERN Theory Division, CH-1211, Switzerland}\\
$^{b}$
{\small Theoretical Physics Division, Ioannina University, Ioannina,
Greece}\\
\end{tabular}
\end{center}

\vspace*{0.15cm}

\begin{center}
{\bf ABSTRACT}
\end{center}

\small{
We review previous work on
neutrino mass textures in gauge theories.
Such textures may arise as a result of flavour
symmetries. In a given theory, it is possible to
give a classification of heavy Majorana neutrino mass
matrices,  assuming
the Dirac masses of the neutrinos to be of the same form
as the ones of the up-quarks.
Heavy Majorana neutrino mass matrices
leading to large neutrino mixing,
are tabulated as an example.
Such solutions are now of interest,
in the light of the Super-Kamiokande data.
}

\section{Introduction}

Although the Standard Model successfully describes
the strong and
electroweak  phenomena, there are still various open
questions, mainly related
to the origin of fermion masses and mixing angles.
An obvious way to explain the observed hierarchies
 that we see at low energies,
would be to assume that some symmetry
(additional to that of the Standard Model)
 is responsible for these patterns.
 An  indication that such 
 additional symmetries are present,
has been the observation that
 the fermion mixing angles and masses have values
 consistent with the appearance of
 ``texture'' zeros in the mass
 matrices \cite{textures,tex2}.

 On the other hand, neutrino data from several experiments
(including 
recent results from Super-Kamiokande\cite{Kam})
seem to require certain
 mixings between various types of massive neutrinos.
 For these,
similar hierarchies are to be expected.
In the framework of the Standard Model, neutrinos are massless,
since both Dirac and Majorana masses are
forbidden by the particle content and the symmetries 
of the theory: A Dirac mass term 
$m (\bar{\nu_{L}}\nu_{R} + \bar{\nu_{R}}\nu_{L})$
does not arise,
due to the absence of right handed neutrinos.
In addition, although 
neutrinos do not carry charge,
a Majorana term $m \nu_{L}^{T} \sigma_{2} 
\nu_{L}$ 
which would violate lepton number
is also forbidden
because it is not $SU(2)$ invariant 
(this term transforms like an 
$I=1$ object under $SU(2)$).
However, in GUTS, the symmetries of the theory allow
the presence of such terms. Moreover, it is 
possible to add (at least) one right handed particle
in the models, thus raising the possibility of a Dirac
mass term as well. If both a Dirac mass and a Majorana mass
(for the isosinglet) neutrino are present, 
the {\em See-Saw} mechanism
may explain the 
lightness of neutrinos relative to the
charged fermions.

The most general neutrino mass matrix for three flavours for both the
isodoublet and isosinglet neutrinos in the current eigenstate basis
\beq
  \chi ^{0}_{R}=(\nu _{R}^{0c},\nu _{R}^0)
\eeq
\beq
  \bar{\chi} ^{0}_{L}=(\bar{\nu} _{L}^{0},\bar{\nu} _{L}^{0c})
\label{eq:n2}
\eeq
takes the form
\begin{equation}
M = \left(
\begin{array}{cc}
M_{L} & D\\
D^{T} & M_{R}
\end{array}
\right),
\label{eq:n3}
\end{equation}
where $M$ a complex symmetric matrix.
The submatrix $M_{L}$ describes the masses
arising in the left-handed sector,
$D$ is the usual Dirac mass matrix
and $M_{R}$ 
contains the entries in the right-handed isosinglet sector.
The mass eigenstates are given by
$\nu _{R}= U^{R} \chi ^{0}_{R}$,
$\nu _{L}= U^{L} \chi ^{0}_{L}$,
$U^{R}$ being a unitary matrix that diagonalizes
$M$ from the right and $U^{L}=(U^{R})^{*}\times exp(i\lambda )$ .
$exp(i\lambda )$ is a diagonal matrix of phases(CP-parities
of the eigenstates. For a real symmetric
matrix the signs of the corresponding eigenvalues) 
\cite{rev}.One rarely deals with the $6 \times 6$ matrix.
In practice one deals with $M_{R}$ amd $m^{eff}_{\nu }$ ('see-saw'
mechanism,see sect.2)

Massive neutrinos have important 
implications for both particle
physics and cosmology. The 
models which admit
neutrinos with the desired properties
are usually severely
constrained, therefore any
experimental evidence about neutrino masses and mixing
will be a powerful probe of physics beyond the standard model.

There are however experimental limits on the masses
and mixing of neutrinos.

\noindent
$\bullet$
Searches for neutrino masses in $\beta$ decays give the limit
$m_{\nu_{e}} \leq 7.3 \,$ eV.

\noindent
$\bullet$
Constraints on the mass of $\nu_{\mu}$ are derived
via the decay process
\[
\pi^{+} \rightarrow \mu^{+} + \nu_{\mu}.
\]
Then, a bound 
$m_{\nu_{\mu}} \leq  170 \, {\rm keV}$
is  derived.

\noindent
$\bullet$
In a similar manner, the $\nu_{\tau}$ is 
constrained by the process
\[
\tau^{-} \rightarrow 2 \pi^{+} 3 \pi^{-} \pi^{0} \nu_{\tau}
\]
to be $\nu_{\tau}
\leq 24 $\, MeV.

\noindent
$\bullet$
An important constraint on Majorana neutrino masses 
for the electron neutrino,
arises from
searches for neutrinoless double $\beta$
decay \cite{rev}
\[
d + d \rightarrow u + u + e + e.
\] 
This process involves violation 
of lepton number by two units.The most popular scenario involves 
intermediate Majorana 
neutrinos (where the particle is the same
as its antiparticle). It can also occur via many
other mechanisms(intermediate exotic scalars,R-parity violating
interactions etc).These other processes ,however, necessarily imply
that the neutrino is a majorana particle.So observation of this
process will show that the neutrinos are majorana particles. No signal
has been seen so far,
which constrains 
the magnitude of a possible mass term to 
be less than a few $eV$.

\noindent
$\bullet$
If neutrinos are massive,
the current eigenstates, $\nu_{i}$,
can be written as
superpositions of the physical 
particles, $\nu_{a}$
\begin{equation}
\nu_{i}(t) = \sum_{a} U_{ia} \nu_{a},
\label{eq:ii}
\end{equation}
where $\nu_{i}$ are the current and $\nu_{a}$ the
mass eigenstates respectively and $U_{ia}$ is
a unitary matrix which diagonalizes the neutrino
mass matrix. 
The evolution equation
of a purely current eigenstate at $t=0$ is then
\begin{equation}
\nu_{i}(t) = \sum_{a} e^{-i E_{a} t } U_{ia} \nu_{a},
\label{evol}
\end{equation}
If the neutrino masses
are not equal, neither are the energies $E_{a}$.
This implies that at a 
subsequent
time, $\nu_{i}(t)$ will be a 
new superposition of the physical
eigenstates than this of eq.~\ref{evol} 
and that there is a 
non-zero probability of finding an
eigenstate $\nu{_i}'$ 
in a beam which was purely $\nu_{i}$.
In this framework, accelerator experiments 
provide strong
bounds on the mixing between $\nu_{e}$
and $\nu_{\mu} $ \cite{em}, 
as well as between $\nu_{\tau}$ and $\nu_{\mu}$ \cite{tm}.

\noindent
$\bullet$
Finally, the LEP experiment on the $Z^{0}$ width 
\cite{LepZ}
limits the number of
light ($m \leq (M_{Z^{0}}/2)$) 
neutrino doublet fields to three.
This bound is also restrictive for other
fields that may couple to the $Z^{0}$, 
such as doublet and triplet
Majorons.

If the neutrinos are light, neutrino oscillation experiments are
the best candidates to measure the small mass differences
$\delta {m^2}$ (from $1eV^2$  down to $10^{-10} eV^2)$.
Moreover, neutrino oscillations may explain the solar neutrino
problem.
The mechanism of neutrino oscillations is only
effective for large neutrino mixing.
However, due to the high density conditions
encountered in the sun, the oscillation can be amplified by
the MSW effect \cite{solar}.
The data from the solar neutrino experiments can thus be described either
by assuming resonant transitions (MSW-effect) or vacuum oscillations.
These require the following ranges for masses and
mixing angles:

a) The small mixing angle solution for the MSW effect requires
\begin{eqnarray}
\delta m^2_{\nu_e\nu_{\alpha}}&\approx &(0.6-1.2)\times 10^{-5}
\; {\rm eV^2} \label{eq:m1}\\
sin^22\theta_{\alpha e}&\approx &(0.6-1.4)\times 10^{-2}\ .
\label{msw}
\end{eqnarray}

b) Vacuum oscillations can solve the solar neutrino puzzle if
\begin{eqnarray}
\delta m^2_{\nu_e\nu_{\alpha}}&\approx &(0.5-1.1)\times 10^{-10}
\; {\rm eV^2}\\
sin^22\theta_{\alpha e}&\ge  &0.75\ ,
\end{eqnarray}
where $\alpha$ is $\mu$ or $\tau$.
The most natural solution  in unified models is obtained through
the MSW-- mass and mixing angle ranges. This solution in particular
requires a light neutrino Majorana mass of the order
\beq
m_{\odot}\approx \sqrt{\delta m^2}\approx 3.0\times 10^{-3}eV\ ,
\eeq
as already given in (\ref{eq:m1}).
Such ultra light masses can be generated effectively in GUT's \cite{LAN}
and SUSY -- GUT's \cite{ROS} by the well known `see--saw' mechanism
\cite{HRR}.

In addition, the atmospheric neutrino problem can be resolved,
in the presence of a large mixing and a small mass splitting involving
the muon neutrino
\cite{atmo}. In this case, bounds from
accelerator and reactor disappearance
experiments indicate that
for $\nu_{e}-\nu_{\mu}$ or
$\nu_{\tau}-\nu_{\mu}$ oscillations
\beq
\delta m^2_{\nu_{\alpha}\nu_{\mu}} \leq
10^{-2} \; {\rm eV}^{2}
\label{at1}
\eeq
\beq
sin^22\theta_{\mu \alpha} \geq  0.51-0.6
\label{at2}
\eeq
where $\alpha$ stands for $e,\tau$ and in (\ref{at2}) the larger
lower limit for $sin^22\theta_{\mu \alpha}$ refers to $\nu_{\mu}-
\nu_{\tau}$ oscillations.
Finally, neutrinos are possible candidates
for structure formation provided they have a mass of order $O({\rm eV})$.
This value can be consistent 
with the bounds from neutrinoless double beta ($\beta\beta_{0\nu}$) decay.
In terms of the neutrino masses and mixing angles, the relevant
 $\beta\beta_{0\nu}$ measurable quantity
can be written as
\beq
|<{ m}_{\nu_e}>|=|\sum_i^3 (U_{ei}) ^2m_{\nu_i}e^{i\lambda_i}|
 \leq 1 eV\ , \label{eq:beta}
\eeq
where $e^{i \lambda_i}$ is the CP-parity of the $i^{th}$ neutrino,
while $U_{ei}$ are the elements of the unitary transformation
 relating the weak and mass neutrino eigenstates(the masses
$m_{\nu _{i}}$ are positive).
It appears therefore, 
that satisfying all the experimental data 
demands  nearly
degenerate mass eigenstates $m_{\nu_i}\approx m_0$, $i=1,2,3$
\cite{psm}. Indeed, structure
formation in the Universe and the COBE data
requires $\sum_{i}m_i \approx 3$ eV, thus
setting the scale of the
masses.  A solution to the atmospheric and solar neutrino deficits,
requires the involved masses to be very similar. In this case the
$\beta\beta_{0\nu}$ bound may be
respected due to mutual cancellations in (\ref{eq:beta})
by opposite CP--phases $e^{i\lambda_i}$.

Passing to neutrino mixing, we see that it
may occur either purely from the neutrino
sector of the theory, or
by the charged lepton mixing. Indeed, 
in complete analogy to the quark
currents the leptonic mixing matrix is
$ V_{tot} =
V_{\ell}V_{\nu}^{\dagger}$
where $V_{\ell}$ diagonalizes the charged lepton mass matrix, while
$V_{\nu}$ diagonalizes the light neutrino mass matrix.
In the former case the mixing is typically too small
to have any impact on the atmospheric
neutrino problem, but may still account
for the solar neutrino problem.
In the latter case, the mixing may be such as to
account for both deficits. 

In the simplest scheme with only one large mixing angle,
one has two possibilities:

1) The solar neutrino problem is resolved by
$\nu_e-\nu_{\mu}$ oscillations and the atmospheric
neutrino problem by
$\nu_{\mu}-\nu_{\tau}$ oscillations.
For this possibility to be viable,
we need a large mixing
angle in the 2-3 entries.

2)The solar neutrino problem is resolved by
$\nu_{e}-\nu_{\tau}$
oscillations and the atmospheric
neutrino problem by
$\nu_e-\nu_{\mu}$
oscillations.
In this case the large angle should be in the
1-2 entries.

\section{Derivation of textures from $U(1)$ symmetries }

Extensions of the
Standard Model with additional $U(1)$ symmetries
can describe the hierarchy of fermion
masses and mixing angles, including neutrinos. An 
example is the model
proposed by Ibanez and Ross \cite{IR}. The structure of the mass
matrices is determined by a family symmetry, $U(1)_{FD}$.
The need to preserve $SU(2)_L$
invariance requires left-handed up and down quarks (leptons)
to have the same charge. This, plus the additional
requirement of symmetric
matrices, indicates that all quarks (leptons) of the same i-th
generation transform with the same charge $\alpha _i(a_i)$.  The
full anomaly free Abelian group involves an additional family
independent component, $U(1)_{FI}$, and with this freedom
$U(1)_{FD}$ is made traceless without any loss of generality. 
If the light Higgs, $H_{2}$, $H_{1}$, responsible for the up and
down quark masses respectively, have $U(1)$ charge so that only
the (3,3) renormalisable Yukawa coupling to $H_{2}$, $H_{1}$ is
allowed, then only the (3,3) element of the associated mass matrix
will be
non-zero. The remaining entries are generated when the
$U(1)$ symmetry is broken. After this
breaking, the mass matrix acquires its structure. 
Different scales $M_{1}$ and $M_2$ are expected
for the down and up quark mass matrices, which 
couple to different Higgs fields.
The lepton mass matrix is determined in a similar
way and for bottom-tau unification one has the picture
\begin{eqnarray}
\frac{M_u}{m_t}\approx \left(
\begin{array}{ccc}
\epsilon^{\mid 2+6a \mid } &
\epsilon^{\mid 3a \mid } &
\epsilon^{\mid 1+3a\mid }
\\
\epsilon^{\mid 3a \mid } &
\epsilon^{ 2 } &
\epsilon^{ 1 } \\
\epsilon^{\mid 1+3a \mid } &
\epsilon^{1 } & 1
\end{array}
\right), 
\frac{M_d}{m_b}\approx \left (
\begin{array}{ccc}
\bar{\epsilon}^{\mid 2+6a \mid } &
\bar{\epsilon}^{\mid 3a \mid } &
\bar{\epsilon}^{\mid 1+3a \mid } \\
\bar{\epsilon}^{\mid 3a \mid } &
\bar{\epsilon}^{ 2 } &
\bar{\epsilon}^{ 1 } \\
\bar{\epsilon}^{\mid 1+3a \mid } &
\bar{\epsilon}^{1} & 1
\end{array}
\right) \nonumber 
\eea
\bea
\frac{M_\ell}{m_\tau}\approx \left(
\begin{array}{ccc}
\bar{\epsilon}^{\mid 2+6a-2b \mid } &
\bar{\epsilon}^{\mid 3a \mid } &
\bar{\epsilon}^{\mid 1+3a-b \mid } \\
\bar{\epsilon}^{\mid 3a \mid } &
\bar{\epsilon}^{ \mid 2(1-b) \mid } &
\bar{\epsilon}^{ \mid 1 -b \mid} \\
\bar{\epsilon}^{\mid 1+3a-b \mid } &
\bar{\epsilon}^{\mid 1-b \mid} &1
\end{array}
\right) \nonumber 
\label{eq:massu}
\end{eqnarray}
The
choices $\bar{\epsilon} 
\equiv (\frac{<\theta >}{M_1})^{|\alpha_2-
\alpha_1|} \approx 0.23$,
$\epsilon 
\equiv (\frac{<\theta >}{M_2})^{|\alpha_2-\alpha_1|}
\approx \bar{\epsilon}^2$,
$a=\alpha_1/(\alpha_2-\alpha_1)=1$
and $\beta 
\equiv (\alpha_2-a_2)/(\alpha_2-\alpha_1) =1/2,0$
give an excellent phenomenological description.

How are neutrino mass matrices included in such a picture?
The first step is to determine the Dirac and heavy Majorana
mass matrices. 
In a scheme 
where we add three generations
of right-handed neutrinos, the light Majorana neutrino
mass matrix is then given by
\begin{equation}
m^{eff}_{\nu}=(m^D_{\nu})^{T}\cdot (M^M_{\nu_R})^{-1}\cdot m^D_{\nu}
\label{eq:meff}
\end{equation}
$SU(2)_L$ fixes the
$U(1)_{FD}$ charge of the left-handed neutrino states to be the
same as the charged leptons. The left- right- symmetry then fixes
the charges of the right-handed neutrinos to be the
same as the left-handed ones.
Therefore, the neutrino Dirac mass is
\cite{DLLRS}
\begin{equation}
\frac{M^D_{\nu_R}}{m_{\nu_{\tau}}}\approx \left (
\begin{array}{ccc}
{\epsilon}^{\mid 2+6a-2b \mid } &
{\epsilon}^{\mid 3a \mid } &
{\epsilon}^{\mid 1+3a-b \mid } \\
{\epsilon}^{\mid 3a \mid } &
{\epsilon}^{ \mid 2(1-b) \mid } &
{\epsilon}^{ \mid 1 -b \mid} \\
{\epsilon}^{\mid 1+3a-b \mid } &
{\epsilon}^{\mid 1-b \mid} & 1
\end{array}
\right)
\label{eq:nud}
\end{equation}
Of course the mass matrix structure of neutrinos is more
complicated, due to the possibility of Majorana masses for the
right-handed components. These arise
from a term of the form
$\nu_R\nu_R\Sigma$  where $\Sigma$ is a $SU(3)\otimes
SU(2)\otimes U(1)$ invariant Higgs scalar field with $I_W=0$ and
$\nu_R$ is a right-handed neutrino.
The possible choices for the
$\Sigma$ $U(1)_{FD}$ charge
 will give a discrete spectrum of
possible forms for the Majorana mass,
$M^M_{\nu}$ \cite{DLLRS,LLR},
whose expansion parameter is the same as the
one for the down and lepton mass matrices \cite{DLLRS}.
>From this previous work, a single $\Sigma$ field
has been used and the main conclusions were
the following:
(a) in schemes where  $\Sigma = \tilde{\bar{\nu}}_R
\tilde{\bar{\nu}}_R$ 
the heavy Majorana mass scale is typically
$10^{13}-10^{14}GeV$, leading to
light neutrinos  between
$O(4-0.4)eV$ for a top quark of $O(200)GeV$ \cite{DLLRS}.
(b) in a particular scheme \cite{LLR},
the mixing in the (2,3) entries
can be quite large. However: 
(c) in none of the  cases does the light Majorana mass matrix
 have degenerate eigenvalues (d) then, for a single $\Sigma$ 
field, the structure of the heavy Majorana and
Dirac mass matrices results in an even larger mass spread
for the light Majorana neutrino masses

However, in principle
there is no reason why this particular conclusion
in the simplest extension of the Standard Model,
should apply in the case of a more complicated symmetry
or with more than one pair of singlet fields,
$\Sigma ,{\bar\Sigma}$  present in the theory \cite{VVV}.
Since in such a case there are many possible patterns,
instead of making a complete search based on symmetries,
it was easier to
work in the opposite way \cite{VVV}, that is: (i) 
Assume that the 
neutrino Dirac mass matrix is proportional to the $u$-quark
mass matrix. Here, we had looked at the
five realistic pairs of texture
zero patterns for the quark mass matrices found
in \cite{RRR}. For completeness, these
appear in table
\ref{table:maj}.
(ii) Study all possible
Majorana neutrino mass matrices with
the maximal number of zeroes, that
lead to a large mixing angle and small
mass splitting between the neutrinos \footnote{
See also \cite{RB}.}.
(iii) Motivate
these phenomenological solutions from symmetries.

\begin{table}
\centering
\begin{tabular}{|c|c|c|} \hline
Solution & $Y_{u}, m^D_{\nu}$ & $Y_{d}$
\\ \hline
1 & $\left(
\begin{array}{ccc}
0 & \sq \la^{6} & 0 \\
\sq \la^{6} & \la^{4} & 0 \\
0 & 0 & 1
\end{array}
\right)$ &
$\left(
\begin{array}{ccc}
0  & 2 \la^{4}  & 0 \\
2 \la^{4}  & 2 \la^{3} & 4 \la^3 \\
0  &  4 \la^3 & 1
\end{array}
\right)$
\\ \hline

2 & $\left(
\begin{array}{ccc}
0 &  \la^{6} & 0 \\
 \la^{6} & 0 & \la^2 \\
0 & \la^2 & 1
\end{array}
\right)$ &
$\left(
\begin{array}{ccc}
0  & 2 \la^{4}  & 0 \\
2 \la^{4}  & 2 \la^{3} & 2\la^3 \\
0  &  2\la^3 & 1
\end{array}
\right)$
\\ \hline

3 & $\left(
\begin{array}{ccc}
0 & 0 & \sq \la^4 \\
0 & \la^{4} & 0 \\
\sq \la^4 & 0 & 1
\end{array}
\right)$ &
$\left(
\begin{array}{ccc}
0  & 2 \la^{4}  & 0 \\
2 \la^{4}  & 2 \la^{3} & 4 \la^3 \\
0  &  4 \la^3 & 1
\end{array}
\right)$
\\ \hline

4 & $\left(
\begin{array}{ccc}
0 & \sq \la^{6} & 0 \\
\sq \la^{6} & \sqrt{3}\la^{4} & \la^2 \\
0 & \la^2 & 1
\end{array}
\right)$ &
$\left(
\begin{array}{ccc}
0  & 2 \la^{4}  & 0 \\
2 \la^{4}  & 2 \la^{3} & 0 \\
0  &  0 & 1
\end{array}
\right)$
\\ \hline

5 & $\left(
\begin{array}{ccc}
0 & 0 & \la^4 \\
0 & \sq \la^{4} & \la^2/\sq \\
\la^4 & \la^2/\sq & 1
\end{array}
\right)$ &

$\left(
\begin{array}{ccc}

0  & 2 \la^{4}  & 0 \\
2 \la^{4}  & 2 \la^{3} & 0 \\
0  &  0 & 1
\end{array}
\right)$  \\
\hline
\end{tabular}
\caption{
Approximate forms for the symmetric
textures.}
\label{table:maj}
\end{table}

\section{Textures for Heavy Majorana mass matrices}

Using the 
Dirac mass matrices from the previous table,
one may look for 
the structure of the heavy Majorana mass matrices
that lead to $m_{eff}$ with
one large angle to solve
the atmospheric neutrino.
Let us start with an atmospheric neutrino mixing residing in the
2-3 submatrix. 
The mixing then can be parametrised as
\beq
V_{\nu} = \left
(\begin{array}{ccc}
1 & 0 & 0 \\
0 & c_{1} & -s_{1} \\
0 & s_{1} & c_{1}
\end{array}
\right)\
\eeq
Then 
$m^{-1}_{eff} = V_{\nu}m_{eff}^{-1diag}V_{\nu}^T$ will have the form
\beq
m_{eff}^{-1} = \left (
\begin{array}{ccc}
\frac{1}{m_{1}} & 0 & 0 \\
0 & \frac{c^2_1}{m_2}+\frac{s_1^2}{m_3} &
c_1s_1(\frac{1}{m_2}-\frac{1}{m_3})
\\
0 &
c_1s_1(\frac{1}{m_2}-\frac{1}{m_3})
& \frac{c_1^2}{m_3}+\frac{s^2_1}{m_2}
\end{array}
\right)
\;
\equiv
\;
\left (
\begin{array}{ccc}
a & 0 & 0 \\
0 & b & d \\
0 & d & c
\end{array}
\right)\ .
\label{eq:form}
\eeq
where $m_{i}$ are the eigenvalues of
$m_{eff}$.  One sees that for $b=c=0$ and
$a=d$, three degenerate neutrinos and
maximal (2-3) mixing is obtained.
Then, the heavy Majorana mass matrix is
given by  $M_{\nu_R} = m_D^{\dagger}\cdot m_{eff}^{-1}\cdot m_D$
and can be easily found once we know $m_D$.
In many unified models, the Dirac neutrino mass
is predicted to have the same structure as the
up-quark mass.
Ramond, Roberts and Ross tabulated the 
quark textures with the maximal number of zeros,
that are consistent with low energy data \cite{RRR}.
Then, we can derive
the forms of the heavy Majorana mass matrices that
lead to a large mixing, for these solutions \cite{VVV}.
The case of large 2-3 mixing is presented
in tables \ref{table:maj2}
and \ref{table:maj3}.
Textures arising from a large mixing in the 1-2 submatrix
appear in table \ref{table:maj41}. Here $m_{eff}^{-1}$ takes a form
similar to (\ref{eq:form}), where the off diagonal elements, $d$, appear
in the 1-2 submatrix.

\begin{table}
\centering
\begin{tabular}{|c|c|c|} \hline
Solution & $ M_{\nu_R}$ & Comments
\\ \hline
1a & $\left(
\begin{array}{ccc}
0 & 0 & \frac{d}{c}\sq \la^{6}  \\
0 & 2\frac{a}{c} \la^{12} & \f{d}{c}\la^{4}  \\
\f{d}{c}\sq\la^6 & \f{d}{c} \la^4 & 1
\end{array}
\right)$ &
for $b=0$
\\ \hline

1b & $\left(
\begin{array}{ccc}
2\frac{b}{d}\la^8 & \sq \frac{b}{d}\la^6 & \sq \la^{2}  \\
\sq \frac{b}{d}\la^6 & \frac{b}{d} \la^{4}+2\frac{a}{d}\la^8 & 1  \\
\sq\la^2 & 1 & 0
\end{array}
\right)$ &
for $c=0$
\\ \hline

2 & $\left(
\begin{array}{ccc}
0 & \f{d}{c}\la^8 & \frac{d}{c} \la^{6}  \\
\f{d}{c}\la^8 & \la^4 +\frac{a}{c}\la^{12}&
\la^2 [+\frac{d}{c}\la^4] \\
\f{d}{c}\la^6 & \la^2 [+\frac{d}{c}\la^4] & 1 [+2\frac{d}{c}\la^2]
\end{array}
\right)$ &
for $b=0$
\\ \hline

3a & $\left(
\begin{array}{ccc}
0 & \sq \la^4 & 0 \\
\sq \la^4 & \f{b}{d}\la^4 & 1 \\
0 & 1 &  2 \f{a}{d}\la^4
\end{array}
\right)$ & for $c=0$
\\ \hline

3b & $\left(
\begin{array}{ccc}
2\la^8 & \sq \frac{d}{c} \la^8 & \sq \la^4 \\
\sq \frac{d}{c} \la^8 & 0 & \frac{d}{c} \la^4 \\
\sq \la^4 & \frac{d}{c} \la^4 &  1+2 \f{a}{c}\la^8
\end{array}
\right)$ & for $b=0$
\\ \hline

4 & $\left(
\begin{array}{ccc}
0 & \frac{d}{c}\sq\la^8 & \f{d}{c}\sq\la^6 \\
\frac{d}{c}\sq\la^8 & \la^4 [+2\sqrt{3}\frac{d}{c}\la^6]
+2\frac{a}{c}\la^{12}
& \la^2 [+(1+\sqrt{3})\frac{d}{c}\la^4] \\
\f{d}{c}\sq\la^6 & \la^2 [+(1+\sqrt{3})\frac{d}{c}\la^4]
& 1 [+2\frac{d}{c}\la^2]
\end{array}
\right)$ &
for $b=0$
\\ \hline


5 & $\left(
\begin{array}{ccc}
0 & \la^6 & \f{1}{2}\la^4 \\
\la^6 & \sq \la^4 & \f{1+2\sq}{4}\la^2 [+\frac{b}{d}
\frac{1}{\sqrt{2}}\la^4] \\
\f{1}{2}\la^4 & \f{1+2\sq}{4}\la^2 [+\frac{b}{d}
\frac{1}{\sqrt{2}}\la^4] & 1 [+\frac{b}{d}\frac{\la^2}{2\sq}]+
\frac{a}{d}\frac{1}{\sq}\la^6
\end{array}
\right)$ &
for $c=0$ \\
\hline
\end{tabular}
\caption{The texture zero solutions of the Majorana mass
matrices associated with
each of the Dirac mass textures of table 1 with a large
mixing in the 2-3 submatrix. We present here cases where
either $b=0$ or $c=0$.
The non-leading powers are in brackets except for
the terms containing the parameter $a=\frac{1}{m_1}$.}
\label{table:maj2}
\end{table}
\begin{table}
\centering
\begin{tabular}{|c|c|c|} \hline
Solution & $ M_{\nu_R}$ & Comments
\\ \hline
1 & $\left(
\begin{array}{ccc}
0&0&{\sqrt{2}}\,d\,{\lambda^6} \\
0&2\,a\, {\lambda^{12}}&d\,{\lambda^4} \\
{\sqrt{2}}\,d\,{\lambda^6}&d\,{\lambda^4}&0
\end{array}
\right)$  &
for $b=c=0$
\\ \hline

2 & $\left(
\begin{array}{ccc}
0 & d\,{\lambda^8}&d\,{\lambda^6}  \\
d\,{\lambda^8}&a\,{\lambda^{12}}&d\,{\lambda^4} \\
d\,{\lambda^6}&d\,{\lambda^4}&2\,d\,{\lambda^2}
\end{array}
\right)$ &
for $b=c=0$
\\ \hline

3 & $\left(
\begin{array}{ccc}
0&{\sqrt{2}}\,d\,{\lambda^8}&0 \\
{\sqrt{2}}\,d\,{\lambda^8}&0&d\,{\lambda^4}\\
0&d\,{\lambda^4}&2\,a\,{\lambda^8}
\end{array}
\right)$ & for $b=c=0$
\\ \hline

4 & $\left(
\begin{array}{ccc}
0&{\sqrt{2}}\,d\,{\lambda^8}&{\sqrt{2}}\,d\, {\lambda^6} \\
{\sqrt{2}}\,d\,{\lambda^8}&2\,{\sqrt{3}}\,d\,
{\lambda^6} + 2\,a\,{\lambda^{12}}&
    d\,{\lambda^4} + {\sqrt{3}}\,d\,{\lambda^4}\\
{\sqrt{2}}\,d\,{\lambda^6}&d\,{\lambda^4} +
 {\sqrt{3}}\,d\,{\lambda^4}&2\,d\,{\lambda^2}
\end{array}
\right)$ &
for $b=c=0$
\\ \hline

5 & $\left(
\begin{array}{ccc}
0&{\sqrt{2}}\,d\,{\lambda^8}&
{{d\,{\lambda^6}}\over {{\sqrt{2}}}} \\
{\sqrt{2}}\,d\,{\lambda^8}&2\,d\,{\lambda^6}&
    {{d\,{\lambda^4}}\over 2} + {\sqrt{2}}\,d\,{\lambda^4} \\
{{d\,{\lambda^6}}\over {{\sqrt{2}}}}&
    {{d\,{\lambda^4}}\over 2} + {\sqrt{2}}\,d\,{\lambda^4}&
    {\sqrt{2}}\,d\,{\lambda^2} + a\,{\lambda^8}
\end{array}
\right)$ &
for $b=c=0$ \\
\hline
\end{tabular}
\caption{Cases as in table 2, but with $b=c=0$.}
\label{table:maj3}
\end{table}
\begin{table}
\centering
\begin{tabular}{|c|c|c|} \hline
Solution & $ M_{\nu_R}$ & Comments
\\ \hline
1 & $\left(
\begin{array}{ccc}
 0 & 2\frac{d}{c}\la^{12} & 0 \\
 2\frac{d}{c}\la^{12} & 2\sq\frac{d}{c}\la^{10} + 2\frac{a}{c}\la^{12} & 0 \\
 0 & 0 & 1
\end{array}
\right)$  &
for $b=0$
\\ \hline

2 & $\left(
\begin{array}{ccc}
 0 & \frac{d}{c}\la^{12} & 0  \\
 \frac{d}{c}\la^{12} & \la^4 + \frac{a}{c}\la^{12} &
 \la^2 + \frac{d}{c}\la^8 \\
 0 & \la^2 + \frac{d}{c}\la^8 & 1
\end{array}
\right)$ &
for $b=0$
\\ \hline

3 & $\left(
\begin{array}{ccc}
 2\la^8 & 0 & \sq\la^4 \\
 0 & 0 & \sq\frac{d}{c}\la^8\\
 \sq\la^4 & \sq\frac{d}{c}\la^8 & 1 + 2\frac{a}{c}\la^8
\end{array}
\right)$ & for $b=0$
\\ \hline

4 & $\left(
\begin{array}{ccc}
 0 & 2\frac{d}{c}\la^{12} & 0 \\
 2\frac{d}{c}\la^{12} & 2\sqrt{6}\frac{d}{c}\la^{10}+2\frac{a}{c}\la^{12} &
 \la^2 + \sq\frac{d}{c}\la^8 \\
 0 & \la^2 + \sq\frac{d}{c}\la^8 & 1
\end{array}
\right)$ &
for $b=0$
\\ \hline

5 & $\left(
\begin{array}{ccc}
 \la^8 & \frac{\la^6}{\sq} & \la^4 \\
 \frac{\la^6}{\sq} & \frac{\la^4}{2} & \frac{\la^2}{\sq} +
 \sq\frac{d}{c}\la^8 \\
 \la^4 & \frac{\la^2}{\sq} + \sq\frac{d}{c}\la^8 &
 1 + 2\frac{d}{c}\frac{\la^6}{\sq} + \frac{a}{c}\la^8
\end{array}
\right)$ &
for $b=0$ \\
\hline
\end{tabular}
\caption{The texture zero solutions of the Majorana mass matrices
associated with each of the Dirac mass textures of table 1 with a large
mixing in the 1-2 submatrix, for the examples with $b=0$. Only cases
for $b=0$ emerge and the solutions for $a=b=0$ follow immediately.}
\label{table:maj41}
\end{table}

Let us  now pass to a discussion of the phenomenology induced by the
forms of $m_{eff}^{-1}$ that have been quoted. We investigate the
case of a large mixing in the 2-3 submatrix. The case of
large mixing in the 1-2 submatrix is very much the same
and leads to analogous conclusions.

There are  two possibilities for texture zero solutions:
$b=0$ or $c=0$\footnote{The case $b=c=0$ e.g. has been already discussed
for the Dirac mass matrix pattern 3 of table 1
in section 4 and implies $\xi=1$,
is therefore in accordance with (\ref{eq:grenzen}).}
that follow.

\noindent (i) {\it c = 0}

This constraint suggests a
 rewriting in terms of the parameter $\xi = -\frac{m_2}{m_3} > 0$.
Then
\beq
c_1 = \frac{1}{\sqrt{1+\xi}}, \; \; \; \; \; \;
s_1 = \frac{\sqrt{\xi}}{\sqrt{1+\xi}}\ ,
\label{eq:sin}
\eeq
\beq
m_{eff}^{-1} =
\left (
\begin{array}{ccc}
\frac{1}{m_1} & 0   &    0 \\
0 & \frac{1-\xi}{m_2} & \frac{\sqrt{\xi}}{m_2} \\
0 & \frac{\sqrt{\xi}}{m_2} &  0
\end{array} \right)
\label{eq:gros}
\eeq
and thus
\beq
sin^2 2\theta_1 = \frac{4\xi}{(1+\xi)^2}\ .
\label{eq:wink}
\eeq
In the neutrino mixing matrix, we add a large angle,
as well as a small one, $\theta_e$,  to explain the solar neutrino problem
\footnote{Renormalisation group effects cause small
deviations from the textures we derive at the GUT scale,
thus small mass differences between the neutrinos
(needed for solving the solar and atmospheric
neutrino problems)
are created in a natural way from the above textures.}.
The neutrino oscillation probabilities are given in terms
of the mixing matrix 
\beq
V_{tot} = V_{e}^{\dagger} V_{\nu} =
\left (
\begin{array}{ccc}
c_{e} & -s_{e} & 0 \\
s_{e} & c_{e} & 0 \\
0 & 0 & 1
\end{array} \right) \cdot
\left (
\begin{array}{ccc}
1 & 0 & 0 \\
0 & c_{1} & s_{1} \\
0 & -s_{1} & c_{1} \\
\end{array}
\right)
\label{eq:multi}
\eeq
\beq
V_{tot} =
\left (
\begin{array}{ccc}
c_{e} & -s_{e}c_1 & -s_{e}s_1 \\
s_{e} & c_{e}c_1 & c_{e}s_1 \\
0 & -s_1 & c_1
\end{array} \right) \ , \label{eq:mixer}
\eeq
where we take
\beq
s_{e} \approx
\sqrt{\frac{m_{e}}{m_{\mu}}} \approx 0.07, \; \; \; c_{e} \approx 1\ .
\label{eq:zahl}
\eeq
Such an ansatz for the charged leptons is most commonly used
and is suggested by the data.
A more general ansatz is definitely more difficult
to handle.

It is now straightforward to calculate the oscillations
$P(\nu_{\alpha} \rightarrow \nu_{\beta})$:
\beq
P(\nu_{\mu} \rightarrow \nu_{\tau}) =
c_e^2 \f{4\xi}{(1+\xi)^2} \sin^2
\f{m_2^2 (1/\xi^2-1)x}{4 E_{\nu}}
\eeq
\beq
P(\nu_{e} \rightarrow \nu_{\tau}) =
s_e^2 \f{4\xi}{(1+\xi)^2} \sin^2
\f{m_2^2 (1/\xi^2-1)x}{4 E_{\nu}}
\eeq
\begin{displaymath}
P(\nu_{e} \rightarrow \nu_{\mu}) =
\sin^2 2\theta_{e} \left[
\f{1}{(1+\xi)} \sin^2
\f{(m_2^2 -m_1^2) x} {4 E_{\nu}}
\right.
\end{displaymath}
\beq
+ \f{\xi}{(1+\xi)} \sin^2
\f{(m_3^2 -m_1^2) x}{4 E_{\nu}}
- \f{\xi}{(1+\xi)^2} \sin^2
\left. \f{(m_3^2 -m_2^2) x}{4 E_{\nu}}
\right] \ .
\label{prob}
\eeq

\noindent (ii) {\it  b = 0}

In this case we have 
\beq
c_1 = \frac{\sqrt{\xi}}{\sqrt{1+\xi}}, \; \; \; \; \; \;
s_1 = \frac{1}{\sqrt{1+\xi}}\ ,
\eeq
\beq
m_{eff}^{-1} =
\left (
\begin{array}{ccc}
\frac{1}{m_1} & 0   &    0 \\
0 & 0 & \frac{\sqrt{\xi}}{m_2} \\
0 & \frac{\sqrt{\xi}}{m_2} & \frac{1-\xi}{m_2}
\end{array} \right)
\eeq
and again the expression (\ref{eq:wink}) for $sin^22\theta_1$.
The oscillation probabilities for $\nu_{\mu} \rightarrow \nu_{\tau}$ and
$\nu_{e} \rightarrow \nu_{\tau}$ remain the same.
For the oscillation $\nu_{e} \rightarrow \nu_{\mu}$ we have to
substitute $\xi \rightarrow \frac{1}{\xi}$.

One may now compare these two possibilities for textures with the
data.
The atmospheric neutrino data implies via (\ref{at1})
that $\xi$ in between
\beq
\xi_1 = 0.23 \; \; {\rm and} \; \; \xi_2 = 4.4\ . \label{eq:grenzen}
\eeq
Given that $\xi_1\xi_2 =1$, the value of $\xi$ selected
merely determines which of the
neutrino masses is heavier, as well as the magnitude of
the masses. Indeed,
from  $m_3^2=\frac{\delta m^2}{1-\xi^2}$
and $m_2^2=m_3^2-\delta m^2$, we observe that,
for a value
$\delta m^2 \approx 0.01$ eV$^2$ as implied by the atmospheric
neutrino data only values of $\xi$ very near unity would give
neutrino masses of order $O(1)$ eV. In particular, one may see that
\beq
m_3 \approx m_2 \approx  1 \; {\rm eV}, \; \; \; {\rm for}\ \; \xi = 0.995\ .
\eeq

After accommodating the atmospheric neutrino data, one can
turn to the discussion of the solar neutrino numbers,
and in this example we interpret them as
$\ne \rightarrow \nm$ oscillations.
{}From (\ref{prob}) we may obtain an effective
$\sin^{2} 2 \theta_{e \mu}$. Depending on the size of $\xi$, the
$\frac{1}{1+\xi}$ or $\frac{\xi}{1+\xi}$ term dominates.
\beq
\xi \ll 1 :\ \  \sin^2 2\theta_{e\mu} \approx \sin^22\theta_{e }
\frac{1}{1+\xi} \approx 1.6 \cdot 10^{-2} \ ,
\eeq
\beq
\xi \gg 1 :\ \  \sin^2 2\theta_{e\mu} \approx \sin^22\theta_{e }
\frac{\xi}{1+\xi} \approx 1.6 \cdot 10^{-2} \ ,
\eeq
when inserting the values of $\theta_e$ and $\xi$.
This is just in agreement with the MSW solution (\ref{msw}).
To satisfy the mass constraints,
$m_1$ must be nearly equal to
$m_2$.
For an average mass $m_0 \approx 1$ eV,
$\delta m_{12}^2 \approx 2m_0 \mid m_2-m_1\mid \approx 10^{-5} eV^2$
indicates the need for a very big degeneracy.
Such a high degree of degeneracy is extremely hard to explain from an
underlying
theory without fine tuning, unless the masses are forced to such values
by symmetries,
as we are going to discuss shortly.

What about neutrinoless double $\beta$-decay and the
COBE data?
For the first one, we obtain
\beq
|<m_{\nu_e}>| = |c_e^2 m_1 + e^{i(\lambda_2 - \lambda_1)}
s_e^2 \left (c_1^2 - \frac{s_1^2}{\xi}e^{i(\lambda_3-\lambda_2)}
\right)m_2 |\ ,
\eeq
where $e^{i(\lambda_2 - \lambda_1)}$
is the relative CP eigenvalue of
$\nu_1$ and $\nu_2$ (the masses here are positive).
Taking $\nu_2$ and $\nu_3$ to have the same CP eigenvalues (as already
discussed in section 2), we obtain
\beq
|<m_{\nu_e}>| = |c_e^2 m_1 + e^{i(\lambda_2 - \lambda_1)}
s_e^2 \left (c_1^2 - \frac{s_1^2}{\xi}\right)m_2 |\ .
\eeq
Now we may again study the texture zeroes.
With (\ref{eq:sin}) we get
\beq
|<m_{\nu_e}>| = c_e^2 m_1 \approx m_1 = O(1) \; {\rm eV}
\eeq
which is consistent with the bound (\ref{eq:beta}).
The above
predictions are consistent with the COBE
data, as well,  since the sum of the masses for the parameter
range we indicate,
can be of order a few eV's, as required.
An identical situation occurs
when the large mixing which explains the atmospheric neutrino
deficit is in the $1-2$ entries of the neutrino mass matrices.

Let us now see how
the quoted mass matrices may arise due to symmetries.
The model of \cite{IR},
which correlates a texture
zero in the (1,3) position
with a texture zero in the (1,1) position, is consistent
with solutions 1, 2 and 4
of \cite{RRR}.
However, the structures 3 and 5 can also arise
from realistic flavour symmetries
\cite{VVV, AKLL}. Indeed, 
in specific GUT groups, the appearance of
zero Clebsch coefficients  is to be expected and
in \cite{AKLL} we derived the zero texture structures for the 
Pati-Salam gauge group combined with a $U(1)$ flavour
symmetry. The same is true
in the presence of residual $Z_2$ symmetries,
and non-trivial mixing in the Higgs-sector \cite{VVV}.
What about the Heavy Majorana mass matrices?
The structures we need require the presence of more
than one singlet fields of the $\Sigma$ type; however
this is not a problem, as in realistic models
many singlets with different $U(1)$ quantum numbers
are expected to appear.
Let us then give an example of how 
the heavy Majorana mass matrices that we presented in
the tables, may arise from a single $U(1)$ symmetry.
Assume the existence of a $\Sigma$ field with
a charge $-1$ (which
makes the (2,3) entry unity), 
as well as the existence of a  second field
$\Sigma'$, with charge $+2$.
Then, the dominant element in the mass matrix 
will be the one with the biggest absolute power in $\bar{\epsilon}$.
In particular, 
the elements (2,2), (2,3) and (3,3) still would couple to
$\Sigma_{1}$ with charge $-1$, while the (1,2)  and (1,3)
will couple to $\Sigma_2$. Then the matrix will be of the form
\begin{eqnarray}
M_{\nu_R}/M_{N}= \left(
\begin{array}{ccc}
0 & \bar{\epsilon} & \bar{\epsilon}^{2} \\
\bar{\epsilon} & \bar{\epsilon} & 1 \\
\bar{\epsilon}^{2} & 1 & \bar{\epsilon}
\end{array}
\right)
\label{eq:mumumu}
\end{eqnarray}
where
\beq
M_{N}=m^2_{t}d \lambda ^4=m^2_{t} d \bar{\epsilon }/\sqrt 2
\eeq
where $m_t$ is the top quark mass.
\beq
M_{N}=1.5 \times 10^{13}
\eeq
The structure of the matrix would be that of the example 3a in Table 3.
This, as we see, is in fact the solution with only
$c=0$ (where the (2,2) element is of order $\bar{\epsilon}$,
and $\bar{\epsilon} = \lambda^4/ \sqrt 2$).\\
>From neutrinoless double beta decay one can extract the limit
\cite{PSVF}
\beq
|< 1/M_{N_e}>|^{-1} \geq 1.6 \times 10^5
\eeq
This arises when the heavy intermediate Majorana neutrinos become
dominant.By diagonalizing the matrix $M_{\nu _{R}}$ we find 
\beq
|< 1/M_{N_e}>|=|\sum_i^3 (U_{ei}) ^2(1/M_{N_i})e^{i\lambda_i}|
\eeq
where $M_{N_i}$ are the heavy eigenmasses. Using 
$\bar{\epsilon}=.23 $ we find 
\beq
| < 1/M_{N_e}>| = 1/{.0118 M_{N}}
\eeq
which leads to
\beq
|<1/M_{N_{e}}>|^{-1} = 1.6 \times 10^{11}
\eeq
which is consistent with the above bound and shows that in our model
the heavy right handed neutrino contribution is unobservable 
in neutrinoless double beta decay.

\section{Summary}

It is possible to derive simple Majorana mass
matrices of right-handed neutrinos, which may explain 
the neutrino experimental data (atmospheric neutrino oscillations,
solar neutrino oscillations in the MSW approach, neutrinoless
double $\beta$-decay and the COBE data).
This requires the existence
of  a right handed
neutrino Majorana mass matrix $M_{\nu_R}$ with a
scale $(10^{12}-10^{13})$ GeV. The solution of the
atmospheric neutrino puzzle resides in a large mixing stemming from
the neutrino mass matrix. Some type of unification or
partial unification implying $m_{\nu}^D \sim m_u$
was adopted.
Along these lines, one can make a classification of
heavy Majorana mass matrices, that for a particular
type of neutrino (and up-quark)
Dirac mass matrix are consistent with the data.


\begin{thebibliography}{99}

\bibitem{textures}
The literature on the subject is vast. Some of the many references are:
H. Fritzsch, Phys. Lett. { 70B} (1977) 436;
{ B73} (1978) 317;
 Nucl. Phys. { B155} (1979) 189;
C. D. Froggatt and H. B. Nilsen, \NPB{147}{79}277;
J. Harvey, P. Ramond and D. Reiss,
\PLB{92}{80}{309};
C. Wetterich, Nucl. Phys. { B261} (1985) 461;
 P. Kaus and S. Meshkov, Mod. Phys. Lett. { A3}
(1988) 1251;
F.J. Gilman and Y. Nir, \ARNP{40}{90}{213};
S. Dimopoulos, L. J. Hall and S. Raby,
\PRL{68}{92}{1984};
\PRD{45}{92}{4195}; H. Arason,
D. J. Casta\~no, P. Ramond and E. J. Piard,
\PRD{47}{93}{232};
G. K. Leontaris and N. D. Tracas, Phys. Lett.{ B303}
(1993) 50;
B.C. Allanach and S.F. King, Nucl.Phys. B456 (1995) 57.


\bibitem{tex2}
Y. Achiman and T. Greiner, Nucl. Phys. { B443} (1995) 3;
P. Binetruy and P. Ramond, Phys. Lett. { B350} (1995) 49;
P. Binetruy, S. Lavignac and P. Ramond,
Nucl. Phys. B477 (1996) 353;
E. Papageorgiu, Z. Phys. { C64} (1994) 509;
Y. Grossman and Y. Nir, Nucl. Phys. { B448} (1995) 30;
B.C. Allanach and S.F. King, Nucl.Phys. B459 (1996) 75.

\bibitem{Kam} Y. Totsuka, International Symbosium on
Lepton Photon Interactions, July 1997, Hamburg.


\bibitem{rev}
For a review, see 
J.D. Vergados, Phys. Rep. { 133} (1986) 1;
S. M. Bilenky and S. T. Petcov, Rev. Mod. Phys. { 59} (1987) 671.
Also: R. N. Mohapatra
and P. B. Pal,
{\em ``Massive neutrinos
in physics and astrophysics''}, World Scientific (1991).



\bibitem{em}
L. A. Ahrens et al., Phys. Rev. D 31, 2732 (1985).

\bibitem{tm}
N. Ushida et al., Phys. Rev. Lett. 57, 2897 (1986).

\bibitem{LepZ}
ALEPH Collab., D. Decamp et al., Phys. Lett. B 235, 399 (1990);
OPAL Collab., M. Z. Akrawy et al., Phys. Lett. B 240, 497 (1990);
DELPHI Collab., P. Abreu et al., Phys. Lett. B 241, 435 (1990);
L3 Collab., B. Adeva et al., Phys. Lett. B 249, 341 (1990).

\bibitem{solar}
See for example,
L. Wolfenstein, Phys.Rev. { D17} (1978) 20;
S. P. Mikheyev and A. Yu Smirnov, Yad. Fiz.{ 42} (1985) 1441;
J. N. Bahcall and W.C. Haxton, Phys.Rev. { D40} (1989) 931;
X. Shi, D. N. Schramm and J. N. Bahcall, Phys.Rev.Lett. {
69} (1992) 717;
P. I. Krastev and S. Petcov, Phys.Lett. { B299} (1993) 94;
 N. Hata and P. Langacker, Phys.Rev. {  D50} (1994) 632
 and references therein.

\bibitem{LAN}
P. Langacker, Phys. Rept. { 72} (1981) 185.

\bibitem{ROS}
G. G. Ross, {\em Grand Unified Theories},
Benjamin Cummings (1985) and references therein.

\bibitem{HRR}
T. Yanagita, Prog. Theor. Phys. B135 (1978) 66;
M. Gell Mann, P. Ramond and R. Slansky,
Supergravity (Eds. P. van Nieuwenhuizen, D. Friedmann, 1979).

\bibitem{atmo}
M. S. Berger et. all, Phys. Lett. { B245}
(1990) 305;
K. S. Hirata et al., Phys. Lett. { B280} (1992) 164;
R. Becker-Szendy et al., Phys. Rev. { D46} (1992) 3720;
Y. Fukuda et al., Phys. Lett. { B335} (1994) 237.

\bibitem{psm}S. Petcov and A. Smirnov, Phys.Lett.
{ B322} (1994) 109.

\bibitem{IR}
L.E. Ib\'a\~nez and G.G. Ross, Phys. Lett. { B332} (1994) 100.

\bibitem{DLLRS}H. Dreiner, G. K. Leontaris, S. Lola,
G. G. Ross and C. Scheich,  Nucl. Phys.{  B346 } (1995) 461.

\bibitem{LLR}
G. Leontaris, S. Lola and G. G. Ross,
Nucl. Phys. { B454} (1995) 25;
G. Leontaris and S. Lola,
hep-ph/9510340, $5^{th}$ Hellenic
School and Workshops on Elementary Particle Physics,
Corfu, September 1995.


\bibitem{VVV}
G. Leontaris, S. Lola,
C. Scheich and J. Vergados,
Phys. Rev. D53 (1996) 6381.


\bibitem{RRR}
P.Ramond, R.G. Roberts and G.G. Ross, Nucl. Phys.
{ B406} (1993) 19.


\bibitem{RB}
P. Binetruy et. al, Nucl. Phys. B496 (1997) 3.

\bibitem{AKLL}
B.C.Allanach, S.F. King, G.K. Leontaris and S. Lola,
Phys. Rev. D56 (1997) 2632;
Phys. Lett. B407 (1997) 275.


\bibitem{PS}
J. Pati and A. Salam, Phys. Rev. D10 (1974) 275.
\bibitem{PSVF}
G.Pantis,F.Simkovic,J.D.Vergados and Amand Faessler,
Phys.Rev. C53 (1997) 695.






\end{thebibliography}
 \end{document}